\documentclass[runningheads]{llncs}
\usepackage{graphicx}
\usepackage{xcolor}

\usepackage{enumitem}
\usepackage{booktabs}
\usepackage{multirow}
\usepackage{subfig}
\usepackage{amsmath}
\usepackage{amssymb}
\usepackage{bm}
\usepackage[ruled,vlined]{algorithm2e}
\usepackage{url}
\usepackage{float}
\usepackage[export]{adjustbox}
\usepackage{marvosym}

\newcommand\blfootnote[1]{%
  \begingroup
  \renewcommand\thefootnote{}\footnote{#1}%
  \addtocounter{footnote}{-1}%
  \endgroup
}

\begin{document}
\title{Robustness of Meta Matrix Factorization Against Strict Privacy Constraints}
\titlerunning{Robustness of MetaMF Against Strict Privacy Constraints}
%

\author{Peter Muellner\inst{1}$^{\textrm{(\Letter)}}$ \and Dominik Kowald \inst{1} \and Elisabeth Lex \inst{2}}

\authorrunning{P. Muellner et al.}
%
\institute{Know-Center GmbH, Graz, Austria \\
\email{\{pmuellner, dkowald\}@know-center.at}\\
\and
Graz University of Technology, Graz, Austria\\
\email{elisabeth.lex@tugraz.at}}
\maketitle              
\begin{abstract}
In this paper, we explore the reproducibility of MetaMF, a meta matrix factorization  framework introduced by Lin et al. MetaMF employs meta learning for federated rating prediction to preserve users' privacy. We reproduce the experiments of Lin et al. on five datasets, i.e., Douban, Hetrec-MovieLens, MovieLens 1M, Ciao, and Jester. Also, we study the impact of meta learning on the accuracy of MetaMF's recommendations. Furthermore, in our work, we acknowledge that users may have different tolerances for revealing information about themselves. Hence, in a second strand of experiments, we investigate the robustness of MetaMF against strict privacy constraints. Our study illustrates that we can reproduce most of Lin et al.'s results. Plus, we provide strong evidence that meta learning is essential for MetaMF's robustness against strict privacy constraints. 

\end{abstract}

\blfootnote{This work has been accepted at ECIR 2021, Reproducibility Track}
\vspace{-1cm}

\section{Introduction}
State-of-the-art recommender systems learn a user model from user and item data and the user's interactions with items to generate personalized recommendations. In that process, however, users' personal information may be exposed, resulting in severe privacy threats. 
As a remedy, recent research makes use of techniques like federated learning~\cite{duriakova2019pdmfrec,chen2020robust,ammad2019federated} or meta learning~\cite{snell2017prototypical,finn2017model} to ensure privacy in recommender systems.
In the federated learning paradigm, no data ever leaves a user's device, and as such, the leakage of their data by other parties is prohibited.
With meta learning, a model gains the ability to form its hypothesis based on a minimal amount of data.

Similar to recent work~\cite{jiang2019improving,chen2018federated}, MetaMF by Lin et al.~\cite{linmeta} combines federated learning with meta learning to provide personalization and privacy. 
Besides, MetaMF exploits collaborative information among users and distributes a private rating prediction model to each user.
Due to MetaMF's recency and its clear focus on increasing privacy for users via a novel framework, we are interested in the reproducibility of Lin et al.'s research.
Additionally, we aim to contribute our own branch of research regarding privacy, i.e., MetaMF's robustness against strict privacy constraints.
This is motivated by a statement of Lin et al. about one critical limitation of MetaMF, i.e., its sensitivity to data scarcity that could arise when users employ strict privacy constraints by withholding a certain amount of their data.
In this regard, every user has a certain privacy budget, i.e., a budget of private data she is willing to share.
Thus, in our paper at hand, the privacy budget is considered a measure of how much data disclosure a user tolerates and is defined as the fraction of rating data she is willing to share with others. 
Thereby, employing small privacy budgets and thus, withholding data, serves as a realization of strict privacy constraints.

Our work addresses MetaMF's limitation against data scarcity and is structured in two parts.
First, we conduct a study with the aim to reproduce the results given in the original work by Lin et al.
Concretely, we investigate two leading research questions, i.e., \emph{RQ1a: How does MetaMF perform on a broad body of datasets?} and \emph{RQ1b: What evidence does MetaMF provide for personalization and collaboration?} 
Second, we present a privacy-focused study, in which we evaluate the impact of MetaMF's meta learning component and test MetaMF's performance on users with different amounts of rating data. 
Here, we investigate two more research questions, i.e., \emph{RQ2a: What is the role of meta learning in the robustness of MetaMF against decreasing privacy budgets?} and \emph{RQ2b: How do limited privacy budgets affect users with different amounts of rating data?}
We address \emph{RQ1a} and \emph{RQ1b} in Section~\ref{sec:reproducibility_study} by testing MetaMF's predictive capabilities on five different datasets, i.e., Douban, Hetrec-MovieLens, MovieLens 1M, Ciao, and Jester. Here, we find that most results provided by Lin et al. can be reproduced. 
In Section~\ref{sec:beyond_reproducibility_study}, we elaborate on \emph{RQ2a} and \emph{RQ2b} by examining MetaMF in the setting of decreasing privacy budgets.
Here, we provide strong evidence of the important role of meta learning in MetaMF's robustness. 
Besides, we find that users with large amounts of rating data are substantially disadvantaged by decreasing privacy budgets compared to users with few rating data.

\section{Methodology}
In this section, we illustrate our methodology of addressing \emph{RQ1a} and \emph{RQ1b}, i.e., the reproducibility of Lin et. al.~\cite{linmeta}, and \emph{RQ2a} and \emph{RQ2b}, i.e., MetaMF's robustness against decreasing privacy budgets.

\subsection{Approach}
\subsubsection{MetaMF.}
Lin et al. recently introduced a novel matrix factorization framework in a federated environment leveraging meta learning.
Their framework comprises three steps. 
First, collaborative information among users is collected and subsequently, utilized to construct a user's collaborative vector.
This collaborative vector serves as basis of the second step.
Here, in detail, the parameters of a private rating prediction model are learned via meta learning.
Plus, in parallel, personalized item embeddings, representing a user's personal ``opinion" about the items, are computed.
Finally, in the third step, the rating of an item is predicted utilizing the previously learned rating prediction model and item embeddings. 
We resort to MetaMF to address \emph{RQ1a}, \emph{RQ1b}, and \emph{RQ2b}, i.e., the reproducibility of results presented by Lin et al. and the influence of decreasing privacy budgets on users with different amounts of rating data.

\subsubsection{NoMetaMF.}
In our privacy-focused study, \emph{RQ2a} addresses the role of meta learning in MetaMF's robustness against decreasing privacy budgets.
Thus, we conduct experiments with and without MetaMF's meta learning component.
For the latter kind of experiments, we introduce NoMetaMF, a variant of MetaMF with no meta learning. 
In MetaMF, a private rating prediction model is generated for each user by leveraging meta learning.
The authors utilize a hypernetwork~\cite{ha2016hypernetworks}, i.e., a neural network, coined meta network, that generates the parameters of another neural network.
Based on the user's collaborative vector $\textbf{c}_u$, the meta network generates the parameters of the rating prediction model, i.e., weights $\textbf{W}^u_l$ and biases $\textbf{b}^u_l$ for layer $l$ and user $u$. This is given by
\begin{align}
    \textbf{h} &= \mathrm{ReLU}(\textbf{W}^*_{h} \textbf{c}_u + \textbf{b}^*_{h}) \label{eq:meta_network_start} \\
    \textbf{W}^u_l &= \textbf{U}^*_{W^u_l} \textbf{h} + \textbf{b}^*_{W^u_l} \\
    \textbf{b}^u_l &= \textbf{U}^*_{b^u_l} \textbf{h} + \textbf{b}^*_{b^u_l} \label{eq:meta_network_end}
\end{align}
where $\textbf{h}$ is the hidden state with the widely-used $\mathrm{ReLU}(x) = \mathrm{max}(0, x)$~\cite{glorot2011deep,hahnloser2000digital} activation function, $\textbf{W}^*_h$, $\textbf{U}^*_{W^u_l}$, $\textbf{U}^*_{b^u_l}$ are the weights and  $\textbf{b}^*_h$, $\textbf{b}^*_{W^u_l}$, $\textbf{b}^*_{b^u_l}$ are the biases of the meta network. 
NoMetaMF excludes meta learning by disabling backpropagation through the meta network in Equations~\ref{eq:meta_network_start}-\ref{eq:meta_network_end}.
Thus, meta parameters $\textbf{W}^*_h$, $\textbf{U}^*_{W^u_l}$, $\textbf{U}^*_{b^u_l}$, $\textbf{b}^*_h$, $\textbf{b}^*_{W^u_l}$, $\textbf{b}^*_{b^u_l}$ will not be learned in NoMetaMF.
While backpropagation is disabled in the meta network, parameters $W^u_l$ and $b^u_l$ are learned over those non-meta parameters in NoMetaMF to obtain the collaborative vector. Hence, the parameters of the rating prediction models are still learned for each user individually, but without meta learning.

Lin et al. also introduce a variant of MetaMF, called MetaMF-SM, which should not be confused with NoMetaMF. 
In contrast to MetaMF, MetaMF-SM does not generate a private rating prediction model for each user individually, but instead utilizes a shared rating prediction model for all users.
Our NoMetaMF model generates an individual rating prediction model for each user but operates without meta learning. Furthermore, we note that in our implementation of NoMetaMF, the item embeddings are generated in the same way as in MetaMF. With NoMetaMF, we aim to investigate the impact of meta learning on the robustness of MetaMF against decreasing privacy budgets, i.e., \emph{RQ2a}.



\subsection{Datasets}
\label{subsec:datasets}
In line with Lin et al., we conduct experiments on four datasets: Douban~\cite{hu2014your}, Hetrec-MovieLens~\cite{cantador2011second}, MovieLens 1M~\cite{harper2015movielens}, and Ciao~\cite{guo2014etaf}.
We observe that none of these datasets comprises a high average number of ratings per item, i.e., 22.6 (Douban), 85.6 (Hetrec-MovieLens), 269.8 (MovieLens 1M), and 2.7 (Ciao).
To increase the diversity of our datasets, we include a fifth dataset to our study, i.e., Jester~\cite{goldberg2001eigentaste} with an average number of ratings per item of 41,363.6.
Furthermore, Lin et al. claimed that several observations about Ciao may be explained by its low average number of ratings per user, i.e., 38.3.
Since Jester exhibits a similarly low average number of ratings per user, i.e., 56.3, we utilize Jester to verify Lin et al.'s claims.
To fit the rating scale of the other datasets, we scale Jester's ratings to a range of [1, 5].
Descriptive statistics of our five datasets are outlined in detail in the following lines.
\emph{Douban} comprises 2,509 users with 893,575 ratings for 39,576 items.
\emph{Hetrec-MovieLens} includes 10,109 items and 855,598 ratings of 2,113 users.
The popular \emph{MovieLens 1M} dataset includes 6,040 users, 3,706 items and 1,000,209 ratings.
\emph{Ciao} represents 105,096 items, with 282,619 ratings from 7,373 users.
Finally, our additional \emph{Jester} dataset comprises 4,136,360 ratings for 100 items from 73,421 users. 

We follow the evaluation protocol of Lin et al. and thus, perform no cross-validation.
Therefore, each dataset is randomly separated into 80\% training set $R_{train}$, 10\% validation set $R_{val}$ and 10\% test set $R_{test}$.
However, we highlight that in the case of Douban, Hetrec-MovieLens, MovieLens 1M, and Ciao, we utilize the training, validation and test set provided by Lin et al.

\subsubsection{Identification of User Groups.}
In \emph{RQ2b}, we study how decreasing privacy budgets influence the recommendation accuracy of user groups with different user behavior. 
That is motivated by recent research~\cite{abdollahpouri2019unfairness,schedl2017distance}, which illustrates differences in recommendation quality for user groups with different characteristics.
As an example, \cite{schedl2017distance} measures a user group's mainstreaminess, i.e., how the user groups' most listened artists match the most listened artists of the entire population. 
The authors split the population into three groups of users with low, medium, and high mainstreaminess, respectively.
Their results suggest that low mainstream users receive far worse recommendations than mainstream users.

In a similar vein, we also split users into three user groups: \emph{Low}, \emph{Med}, and \emph{High}, referring to users with a low, medium, and a high number of ratings, respectively.
To precisely study the effects of decreasing privacy budgets on each user group, we generate them such that the variance of the number of ratings is low, but yet, include a sufficiently large number of users.
For this matter, each of our three user groups includes 5\% of all users.
In detail, we utilize the 5\% of users with the least ratings (i.e, \emph{Low}), the 5\% of users with the most ratings (i.e., \emph{High}) and the 5\% of users, whose number of ratings are the closest to the median (i.e., \emph{Med}).
Thus, each user group consists of 125 (Douban), 106 (Hetrec-MovieLens), 302 (MovieLens 1M), 369 (Ciao), and 3,671 (Jester) users.

\subsection{Recommendation Evaluation}
In concordance to the methodology of Lin et al., we minimize the mean squared error (MSE) between the predicted $\hat{r} \in \hat{R}$ and the real ratings $r \in R$ as the objective function for training the model.
Additionally, we report the MSE and the mean absolute error (MAE) on the test set $R_{test}$ to estimate our models' predictive capabilities.
Since we dedicate parts of this work to shed light on MetaMF's and NoMetaMF's performance in settings with different degrees of privacy, we illustrate how we simulate decreasing privacy budgets and how we evaluate a model's robustness against these privacy constraints.

\subsubsection{Simulating Different Privacy Budgets.}
To simulate the reluctance of users to share their data, we propose a simple sampling procedure in Algorithm~\ref{algo:algorithm}.
Let $\beta$ be the privacy budget, i.e., the fraction of data to be shared.
First, a user $u$ randomly selects a fraction of $\beta$ of her ratings without replacement.
Second, the random selection of ratings $R^\beta_u$ is then shared by adding it to the set $R^\beta$. 
That ensures that (i) each user has the same privacy budget $\beta$ and (ii) each user shares at least one rating to receive recommendations.
The set of shared ratings $R^\beta$ without held back ratings then serves as a training set for our models.
\begin{algorithm}[!htb]
\SetAlgoLined
\SetKwInOut{Input}{Input}
\Input{Ratings $R$, Users $U$ and privacy budget $\beta$.}
\KwResult{Shared ratings $R^\beta$, with a fraction of $\beta$ of each user's ratings.}
    $R^\beta = \{\}$ \\
    \For{$u \in U$}{
    $R^\beta_u = \{R'_u \subseteq R_u: |R'_u| / |R_u| = \beta\}$ \\
    $R^\beta = R^\beta \cup R^\beta_u$
    }
 \caption{Sampling procedure for simulating privacy budget $\beta$.}
 \label{algo:algorithm}
\end{algorithm}

\subsubsection{Measuring Robustness.} 
Our privacy-focused study is concerned with discussing MetaMF's robustness against decreasing privacy budgets.
We quantify a model's robustness by how the model's predictive capabilities change by decreasing privacy budgets.
In detail, we introduce a novel accuracy measurement called $\Delta\mathrm{MAE}@\beta$, which is a simple variant of the mean absolute error. 
\begin{definition}[$\Delta\mathrm{MAE}@\beta$]
    The relative mean absolute error $\Delta\mathrm{MAE}@\beta$ measures the predictive capabilities of a model $M$ under a privacy budget $\beta$ relative to the predictive capabilities of $M$ without any privacy constraints.
    \begin{align}
        \mathrm{MAE}@\beta &= \frac{1}{|R_{test}|}\sum_{r_{u, i} \in R_{test}} |(r_{u, i} - M(R^\beta_{train}, \theta)_{u, i})| \\
        \Delta\mathrm{MAE}@\beta &= \frac{\mathrm{MAE}@\beta}{\mathrm{MAE}@1.0}
    \end{align}
    where $M(R^\beta_{train}, \theta)_{u, i}$ is the estimated rating for user $u$ on item $i$ for $M$ with parameters $\theta$ being trained on the dataset $R^\beta_{train}$ and $|\cdot|$ is the absolute function. Please note that the same $R_{test}$ is utilized for different values of $\beta$.
\end{definition}
Furthermore, it is noteworthy that the magnitude of $\Delta\mathrm{MAE}@\beta$ measurements does not depend on the underlying dataset, as it is a relative measure.
Thus, one can compare a model's $\Delta\mathrm{MAE}@\beta$ measurements among different datasets.

\subsection{Source Code and Materials} 
For the reproducibility study, we utilize and extend the original implementation of MetaMF, which is provided by the authors alongside the Douban, Hetrec-MovieLens, MovieLens 1M, and Ciao dataset samples via BitBucket\footnote[1]{\url{https://bitbucket.org/HeavenDog/metamf/src/master/}, Last accessed Oct. 2020}. 
Furthermore, we publish the entire Python-based implementation of our work on GitHub\footnote[2]{\url{https://github.com/pmuellner/RobustnessOfMetaMF}} and our three user groups for all five datasets on Zenodo\footnote[3]{\url{https://doi.org/10.5281/zenodo.4031011}}~\cite{mullner_peter_2020_4031011}.

We want to highlight that we are not interested in outperforming any state-of-the-art approaches on our five datasets.
Thus, we refrain from conducting any hyperparameter tuning or parameter search and utilize precisely the same parameters, hyperparameters, and optimization algorithms as Lin et al~\cite{linmeta}. 

\section{Reproducibility Study}
\label{sec:reproducibility_study}
In this section, we address \emph{RQ1a} and \emph{RQ1b}. 
As such, we repeat experiments by Lin et al.~\cite{linmeta} to verify the reproducibility of their results.
Therefore, we evaluate MetaMF on the four datasets Douban, Hetrec-MovieLens, MovieLens 1M, and Ciao.
Additionally, we measure its accuracy on the Jester dataset.
Please note that we strictly follow the evaluation procedure as in the work to be reproduced.

\begin{table}[!t]
    \caption{MetaMF's error measurements (reproduced/original) for our five datasets alongside the MAE (mean absolute error) and the MSE (mean squared error) reported in the original paper. The non-reproducibility of the MSE on the Ciao dataset can be explained by the particularities of the MSE and the Ciao dataset. All other measurements can be reproduced (\emph{RQ1a}).}
    \centering
    \begin{tabular}{l l l}
    \toprule
    Dataset & MAE & MSE \\ \midrule
    Douban & 0.588/0.584 & 0.554/0.549 \\
    Hetrec-MovieLens & 0.577/0.571 & 0.587/0.578 \\
    MovieLens 1M & 0.687/0.687 & 0.765/0.760 \\
    Ciao & 0.774/0.774 & 1.125/1.043 \\ \midrule
    Jester & 0.856/- & 1.105/- \\ \bottomrule
    \end{tabular}
    \label{tab:err_reproduction}
\end{table}

We provide MAE (mean absolute error) and MSE (mean squared error) measurements on our five datasets in Table~\ref{tab:err_reproduction}.
It can be observed that we can reproduce the results by Lin et al. up to a margin of error smaller than $2\%$.
Only in the case of the MSE on the Ciao dataset, we obtain different results.
Due to the selection of random batches during training, our model slightly deviates from the one utilized by Lin et al.
Thereby, also, the predictions are likely to differ marginally.
As described in~\cite{willmott2005advantages}, the MSE is much more sensitive to the variance of the observations than the MAE.
Thus, we argue that the non-reproducibility of the MSE on the Ciao dataset can be explained by the sensitivity of the MSE on the variance of the observations in each batch. 
In detail, we observed in Section~\ref{subsec:datasets} that Ciao comprises very few ratings but lots of items.
Thus, the predicted ratings are sensitive to the random selection of training data within each batch.
However, it is noteworthy that we can reproduce the more stable MAE on the Ciao dataset.
Hence, we conclude that our results provide strong evidence of the originally reported measurements being reproducible, enabling us to answer \emph{RQ1a} in the affirmative.

\begin{figure}[!t]
    \centering
    \begin{adjustbox}{minipage=\linewidth,scale=0.88}
    \centering
    \subfloat[t][Douban]{\includegraphics[width=0.33\linewidth]{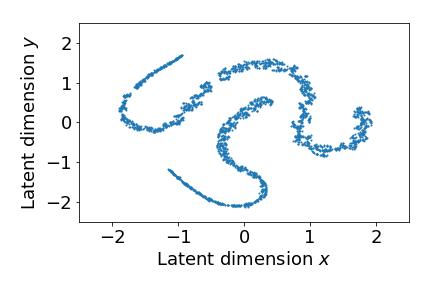}}
    \subfloat[t][Hetrec-MovieLens]{\includegraphics[width=0.33\linewidth]{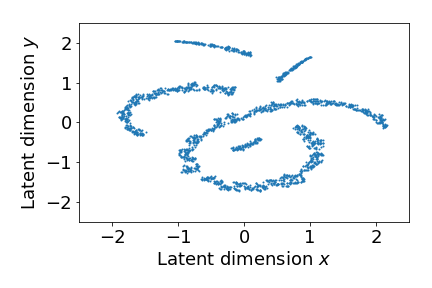}}
    \subfloat[t][MovieLens 1M]{\includegraphics[width=0.33\linewidth]{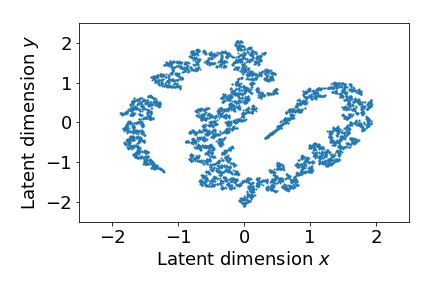}}
    \newline
    \subfloat[t][Ciao]{\includegraphics[width=0.33\linewidth]{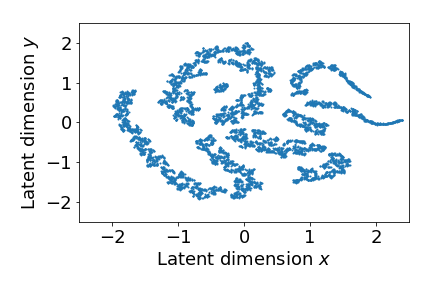}\label{fig:weight_embeddings_ciao}}
    \subfloat[t][Jester]{\includegraphics[width=0.33\linewidth]{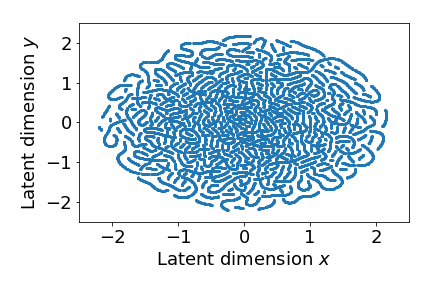}}
    \end{adjustbox}
    \caption{MetaMF's weights embeddings of the first layer of the rating prediction models. One observation corresponds to an individual user (\emph{RQ1b}).}
    \label{fig:weight_embeddings}
\end{figure}

Next, we study the rating prediction models' weights and the learned item embeddings.
Again, we follow the procedure of Lin et al. and utilize the popular t-SNE (t-distributed stochastic neighborhood embedding)~\cite{maaten2008visualizing} method to reduce the dimensionality of the weights and the item embeddings to two dimensions.
Since Lin et al. did not report any parameter values for t-SNE, we rely on the default parameters, i.e., we set the perplexity to 30~\cite{maaten2008visualizing}.
After the dimensionality reduction, we standardize all observations $x \in X$ by $\frac{x-\mu}{\sigma}$, where $\mu$ is the mean and $\sigma$ is the standard deviation of $X$.
The rating prediction model of each user is defined as a two-layer neural network.
However, we observe that Lin et al. did not describe what layer's weights they visualize.
Correspondences with the leading author of Lin et al. clarified that in their work, they only describe the weights of the first layer of the rating prediction models.
The visualizations of the first layer's weights of the rating prediction models on our five datasets are given in Figure~\ref{fig:weight_embeddings}.

In line with Lin et al., we discuss the weights and the item embeddings with respect to personalization and collaboration.
As the authors suggest, personalization leads to distinct weight embeddings and collaboration leads to clusters within the embedding space.
First, we observe that MetaMF tends to generate different weight embeddings for each user.
Second, the visualizations exhibits well-defined clusters, which indicates that MetaMF can exploit collaborative information among users.
However, our visualizations of the weights deviate slightly from the ones reported by Lin et al.
Similar to the reproduction of the accuracy measurements in Table~\ref{tab:err_reproduction}, we attribute this to the inability to derive the exact same model as Lin et al.
Besides, t-SNE comprises random components and thus, generates slightly varying visualizations.
However, the weights for the Ciao dataset in Figure~\ref{fig:weight_embeddings_ciao} illustrate behavior that contradicts Lin et al.'s observations.
In the case of the Ciao dataset, they did not observe any form of clustering and attributed this behavior to the small number of ratings per user in the Ciao dataset. 
To test their claim, we also illustrate the Jester dataset with a similarly low number of ratings per user. 
In contrast, our visualizations indeed show well-defined clusters and different embeddings. 
We note that Jester exhibits many more clusters than the other datasets due to the much larger number of users.
Overall, we find that both, Ciao and Jester, do not support the claim made by Lin et al.
However, we see the possibility that this observation may be caused by randomness during training.

Due to space limitations, we refrain from visualizing the item embeddings.
It is worth noticing that our observations on the weights also hold for the item embeddings.
In detail, our visualizations exhibit indications of collaboration and personalization for all datasets.
Overall, we find the visualizations of the weights and the item embeddings presented by Lin et al. to be reproducible for the Douban, Hetrec-MovieLens, and MovieLens 1M datasets and thus, we can also positively answer \emph{RQ1b}.

\section{Privacy-Focused Study}
\label{sec:beyond_reproducibility_study}
In the following, we present experiments that go beyond reproducing Lin et al.'s work~\cite{linmeta}. 
Concretely, we explore the robustness of MetaMF against decreasing privacy budgets and discuss \emph{RQ2a} and \emph{RQ2b}.
More detailed, we shed light on the effect of decreasing privacy budgets on MetaMF in two settings: (i) the role of MetaMF's meta learning component and (ii) MetaMF's ability to serve users with different amounts of rating data equally well.
\begin{figure}[!htb]
    \centering
    \subfloat[t][MetaMF]{\includegraphics[width=0.49\linewidth]{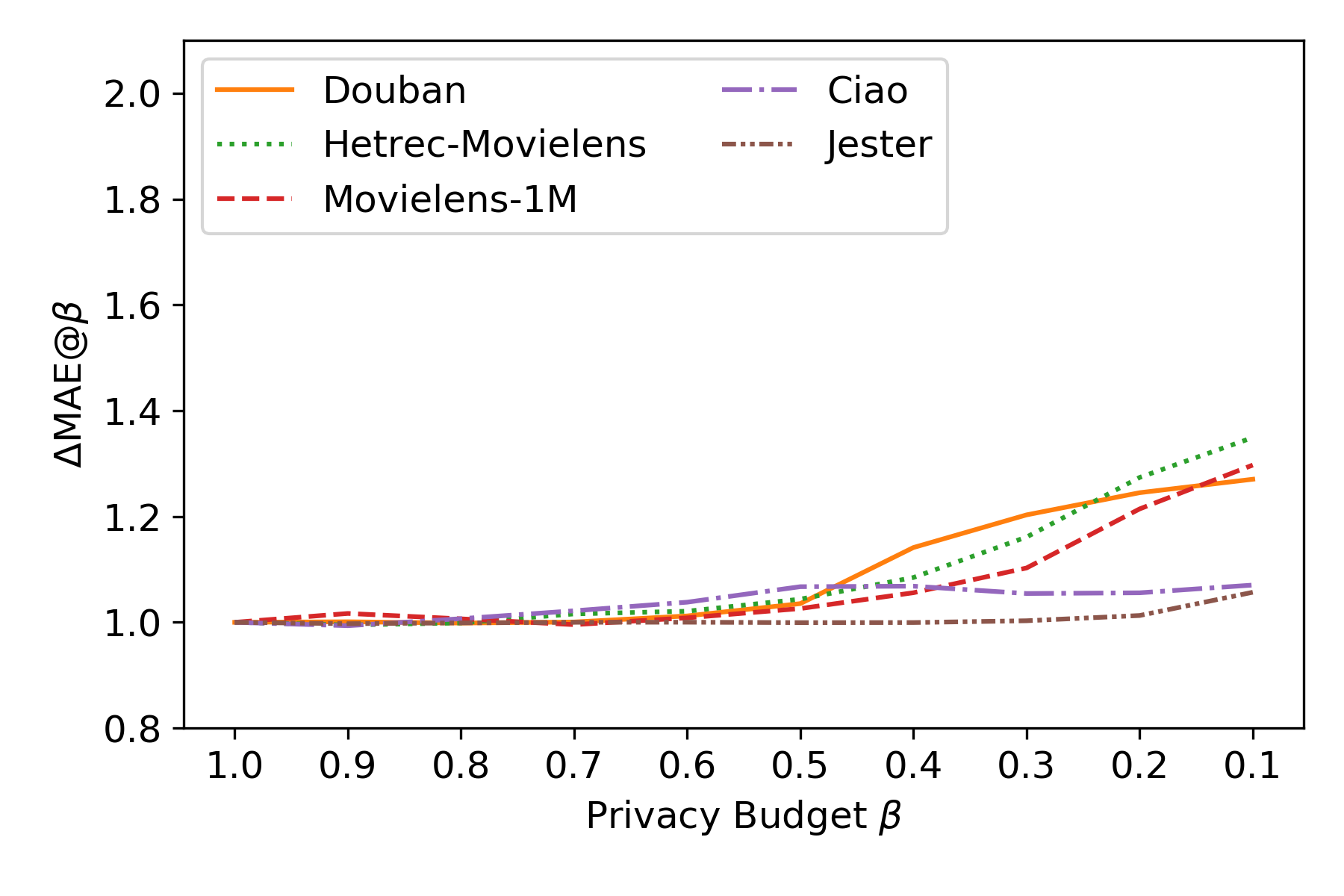}\label{fig:meta_nometa_comparison_a}}
    \subfloat[t][NoMetaMF]{\includegraphics[width=0.49\linewidth]{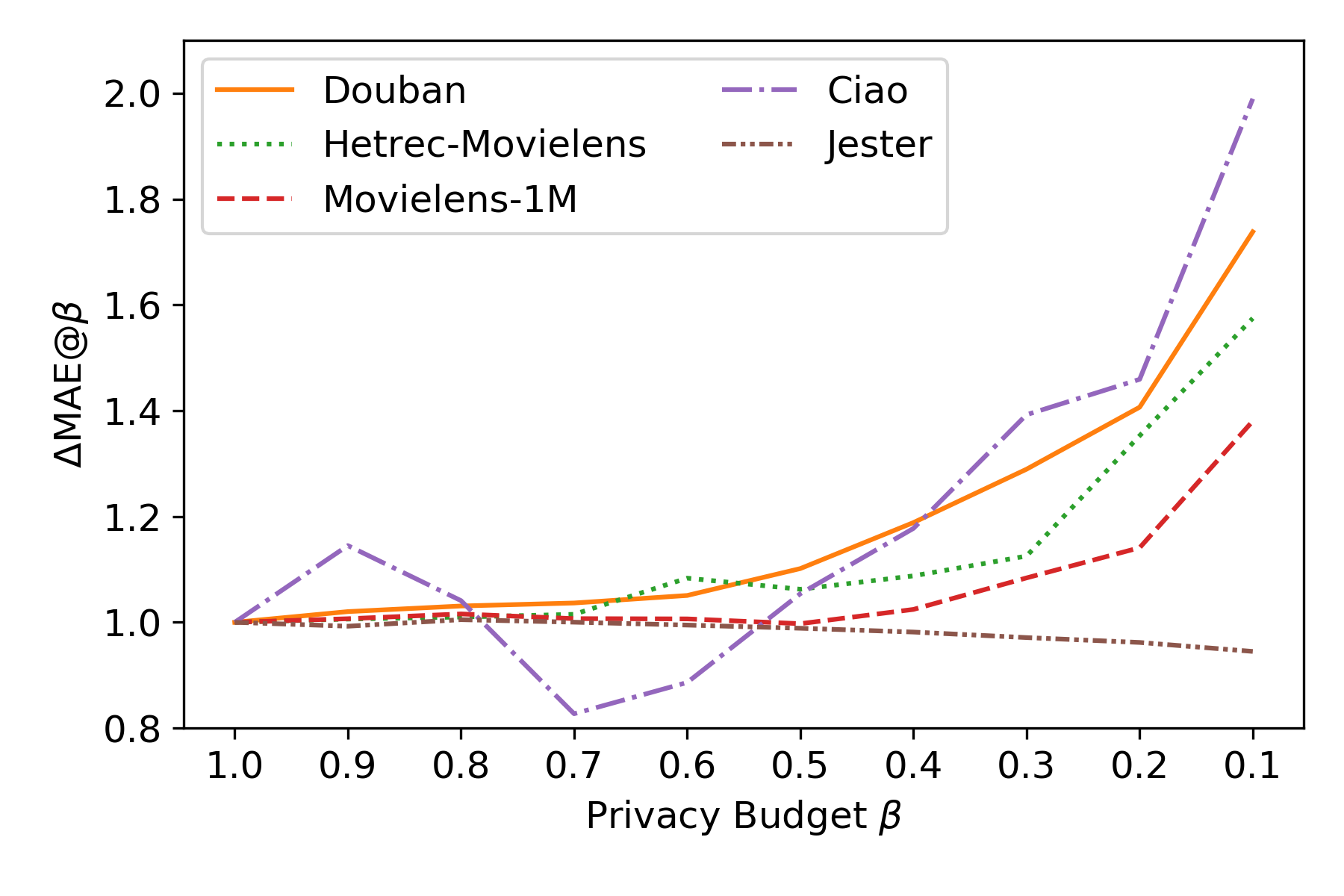}\label{fig:meta_nometa_comparison_b}}
    \caption{$\Delta\mathrm{MAE}@\beta$ measurements on (a) MetaMF and (b) NoMetaMF, in which meta learning is disabled. Especially for small privacy budgets, MetaMF yields a much more stable accuracy than NoMetaMF (\emph{RQ2a}).}
    \label{fig:meta_nometa_comparison}
\end{figure}

First, we compare MetaMF to NoMetaMF in the setting of decreasing privacy budgets.
Therefore, we utilize our sampling procedure in Algorithm~\ref{algo:algorithm} to generate datasets with different privacy budgets.
In detail, we construct 10 training sets, i.e., $\{R^\beta_{train}: \beta \in \{1.0, 0.9, \dots, 0.2, 0.1\}\}$, on which MetaMF and NoMetaMF are trained on.
Then, we evaluate both models on the test set $R_{test}$.
It is worth noticing that $R_{test}$ is the same for all values of $\beta$ to enable a valid comparison.
Our results in Figure~\ref{fig:meta_nometa_comparison_a} illustrate that for all datasets, MetaMF preserves its predictive capabilities well, even with decreasing privacy budgets.
However, a privacy budget of $\approx 50\%$ seems to be a critical threshold.
The $\Delta\mathrm{MAE}@\beta$ only marginally increases for $\beta > 0.5$, but rapidly grows for $\beta \leq 0.5$ in the case of the Douban, Hetrec-MovieLens, and MovieLens 1M dataset.
In other words, a user could afford to withhold $\leq 50\%$ of her data and still get well-suited recommendations.
Additionally, the $\Delta\mathrm{MAE}@\beta$ remains stable for the Ciao and Jester dataset.
Similar observations can be made about the results of NoMetaMF in Figure~\ref{fig:meta_nometa_comparison_b}.
Again, the predictive capabilities remain stable for $\beta > 0.5$ in the case of Douban, Hetrec-MovieLens, and MovieLens 1M, but decrease tremendously for higher levels of privacy.
Our side-by-side comparison of MetaMF and NoMetaMF in Figure~\ref{fig:meta_nometa_comparison} suggests that both methods exhibit robust behavior for large privacy budgets (i.e., $\beta > 0.5$), but exhibit an increasing MAE for less data available (i.e., $\beta \leq 0.5$). 
However, we would like to highlight that the increase of the MAE is much worse for NoMetaMF than for MetaMF.
Here, the $\Delta\mathrm{MAE}@\beta$ indicates that the MAE for NoMetaMF increases much faster than the MAE for MetaMF for decreasing privacy budgets.
This observation pinpoints the importance of meta learning and personalization in settings with a limited amount of data per user, i.e., a high privacy level.
Thus, concerning \emph{RQ2a}, we conclude that MetaMF is indeed more robust against decreasing privacy budgets than NoMetaMF, but yet, requires a sufficient amount of data per user.
\begin{figure}[!ht]
    \centering
    \begin{adjustbox}{minipage=\linewidth,scale=0.88}
    \centering
    \subfloat[t][$\beta=1.0$]{\includegraphics[width=0.33\linewidth]{figures/ml/meta/0p_weights1.png}}
    \subfloat[t][$\beta=0.5$]{\includegraphics[width=0.33\linewidth]{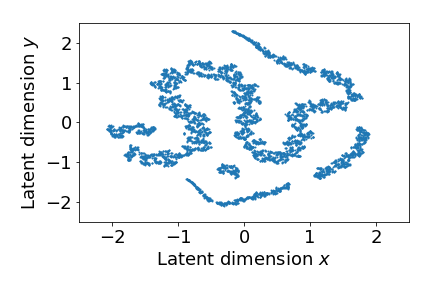}}
    \subfloat[t][$\beta=0.1$]{\includegraphics[width=0.33\linewidth]{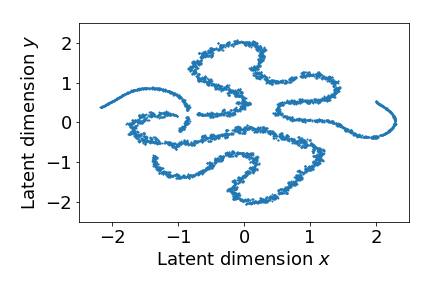}}
    \newline
    \subfloat[t][$\beta=1.0$]{\includegraphics[width=0.33\linewidth]{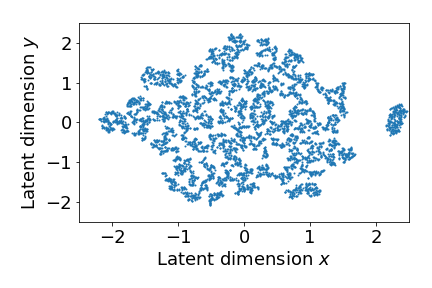}}
    \subfloat[t][$\beta=0.5$]{\includegraphics[width=0.33\linewidth]{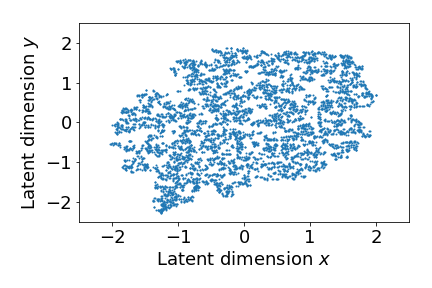}}
    \subfloat[t][$\beta=0.1$]{\includegraphics[width=0.33\linewidth]{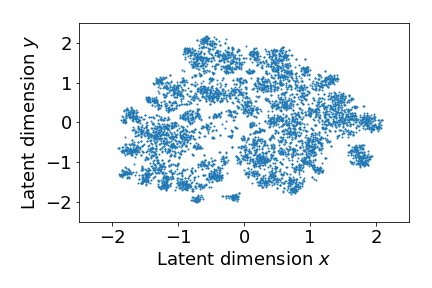}}
    \end{adjustbox}
    \caption{Weights of the first layer of the rating prediction models for the MovieLens 1M dataset. Figures (a), (b), (c) depict MetaMF, whereas Figures (d), (e), (f) depict NoMetaMF, in which meta learning is disabled. No well-defined clusters are visible for NoMetaMF, which indicates the inability to 
    exploit collaborative information among users (\emph{RQ2a}).}
    \label{fig:meta_nometa_embeddings}
\end{figure}

Next, we compare MetaMF to NoMetaMF with respect to their ability for personalization and collaboration in the setting of decreasing privacy budgets.
As explained in Section~\ref{sec:reproducibility_study}, we refer to Lin et al., which suggest that personalization leads to distinct weight embeddings and collaboration leads to clusters within the embedding space.
In Figure~\ref{fig:meta_nometa_embeddings}, we illustrate the weights of the first layer of the rating prediction models of MetaMF and NoMetaMF for the \mbox{MovieLens 1M} dataset for different privacy budgets (i.e., $\beta \in \{1.0, 0.5, 0.1\}$).
Again, we applied t-SNE to reduce the dimensionality to two dimensions, followed by standardization to ease the visualization.
In the case of MetaMF, we observe that it preserves the ability to generate different weights for each user for decreasing privacy budgets.
Similarly, well-defined clusters can be seen, which indicates that MetaMF also preserves the ability to capture collaborative information among users.
In contrast, our visualizations for NoMetaMF do not show well-defined clusters. This indicates that NoMetaMF loses the ability to exploit collaborative information among users.
Due to limited space, we refrain from presenting the weights of the first layer of the rating prediction models for the other datasets.
However, we observe that MetaMF outperforms NoMetaMF in preserving the collaboration ability for decreasing privacy budgets on the remaining four datasets, which is also in line with our previous results regarding \emph{RQ2a}.

In the following, we elaborate on how the high degree of personalization in MetaMF impacts the recommendations of groups of users with different amounts of rating data.
In a preliminary experiment, we measure the MAE on our three user groups \emph{Low}, \emph{Med}, and \emph{High} on our five datasets in Table~\ref{tab:mae_usergroups}.
Except for the Ciao dataset, our results provide evidence that \emph{Low} is served with significantly worse recommendations than \emph{High}.
In other words, users with lots of ratings are advantaged over users with only a few ratings.
\begin{table}[!htb]
    \caption{MetaMF's MAE (mean absolute error) measurements for our three user groups on the five datasets. Here, we simulated a privacy budget of $\beta=1.0$. 
    According to a one-tailed t-Test, \emph{Low} is significantly disadvantaged over \emph{High}, indicated by *, i.e., $\alpha=0.05$ and ****, i.e., $\alpha=0.0001$ (\emph{RQ2b}).}
    \centering
    \begin{tabular}{l l l l}
    \toprule
    Dataset & \emph{Low} & \emph{Med} & \emph{High} \\ \midrule
    Douban* & 0.638 & 0.582 & 0.571 \\
    Hetrec-MovieLens**** & 0.790 & 0.603 & 0.581 \\
    MovieLens 1M**** & 0.770 & 0.706 & 0.673 \\
    Ciao & 0.773 & 0.771 & 0.766 \\ \midrule
    Jester**** & 1.135 &  0.855 & 0.811 \\ \bottomrule
    \end{tabular}
    \label{tab:mae_usergroups}
\end{table}

\begin{figure}[!b]
    \centering
    \begin{adjustbox}{minipage=\linewidth,scale=0.88}
    \centering
    \subfloat[t][Douban]{\includegraphics[width=0.33\linewidth]{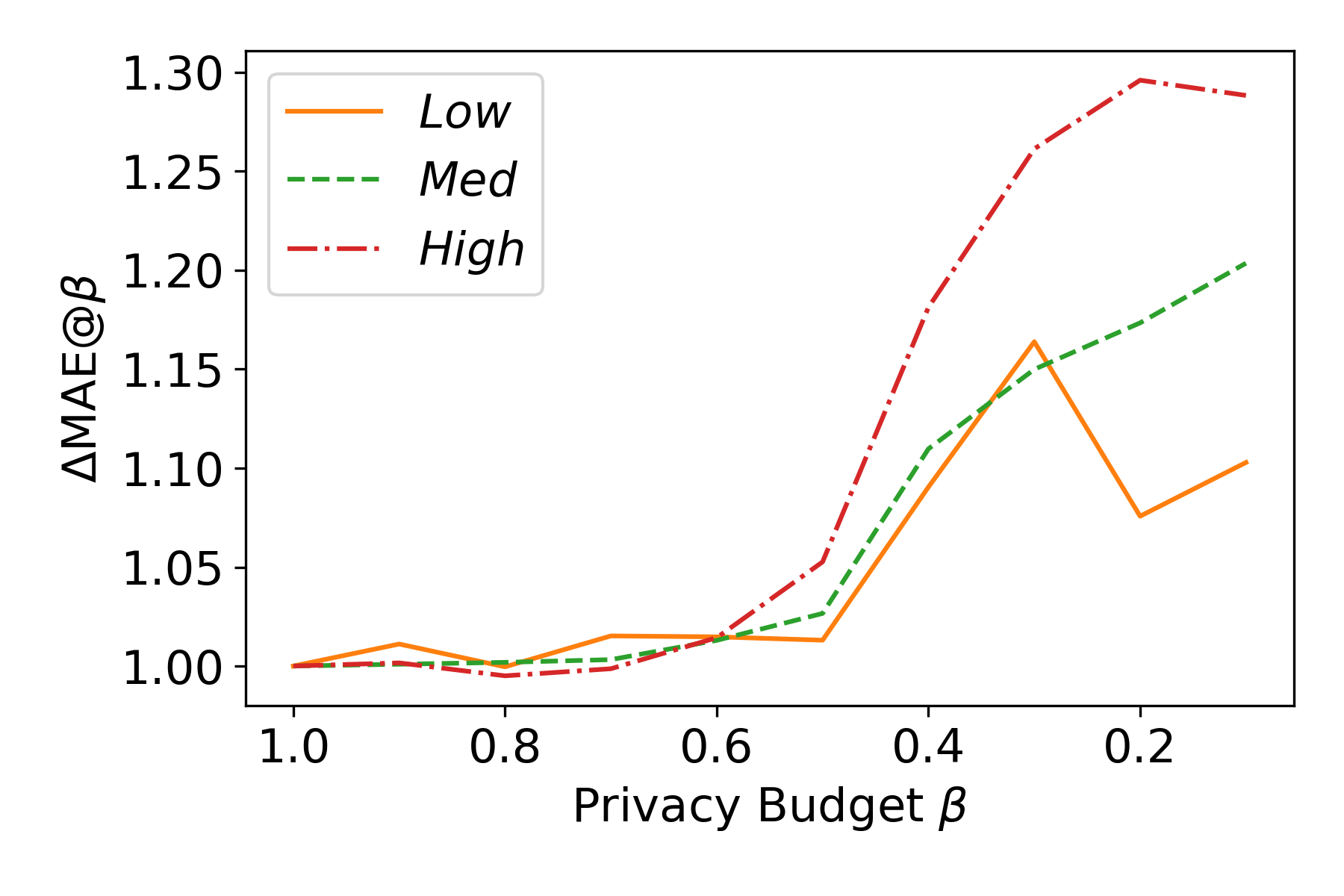}\label{fig:usergroups_a}}
    \subfloat[t][Hetrec-MovieLens]{\includegraphics[width=0.33\linewidth]{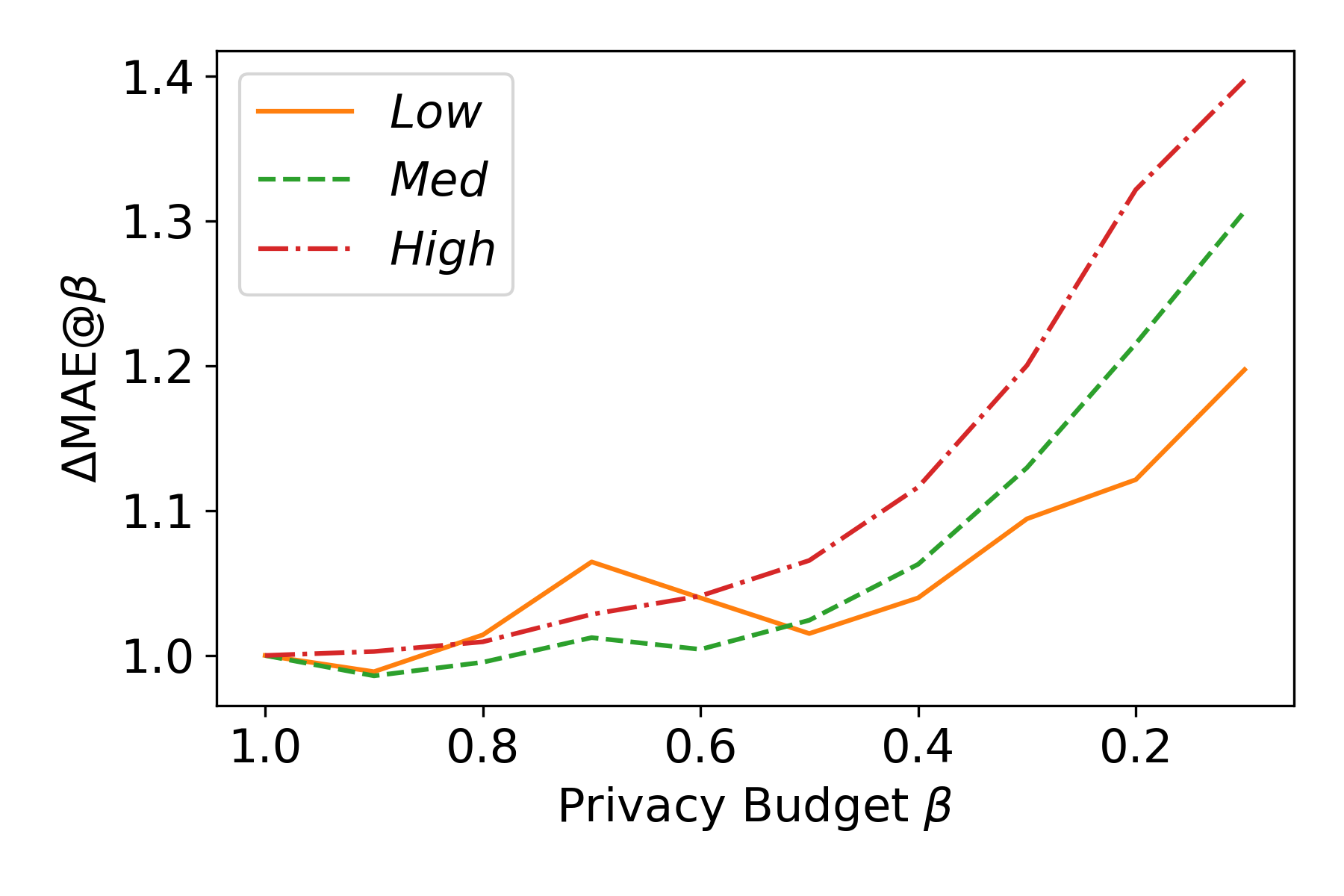}\label{fig:usergroups_b}}
    \subfloat[t][MovieLens 1M]{\includegraphics[width=0.33\linewidth]{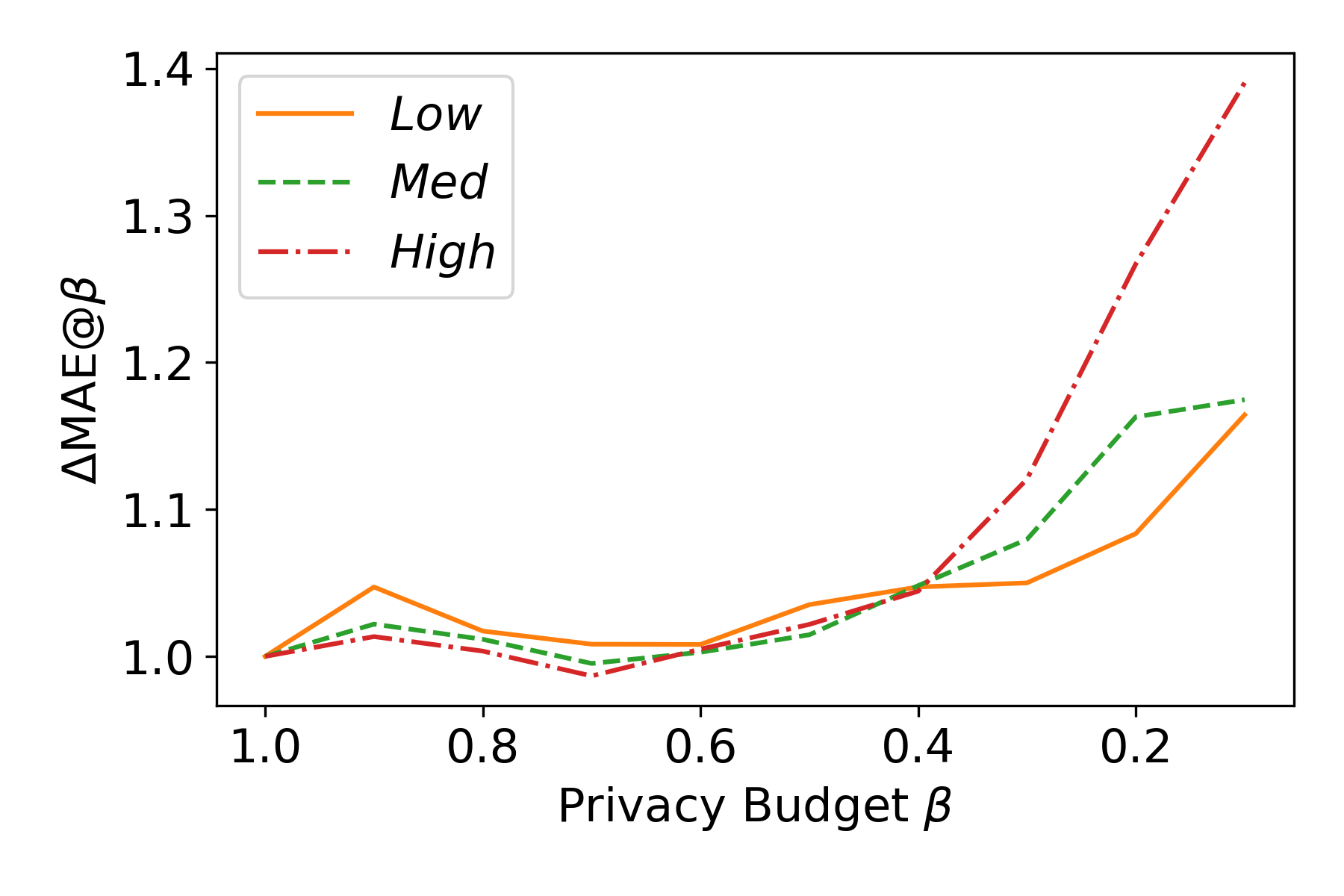}\label{fig:usergroups_c}}
    \newline
    \subfloat[t][Ciao]{\includegraphics[width=0.33\linewidth]{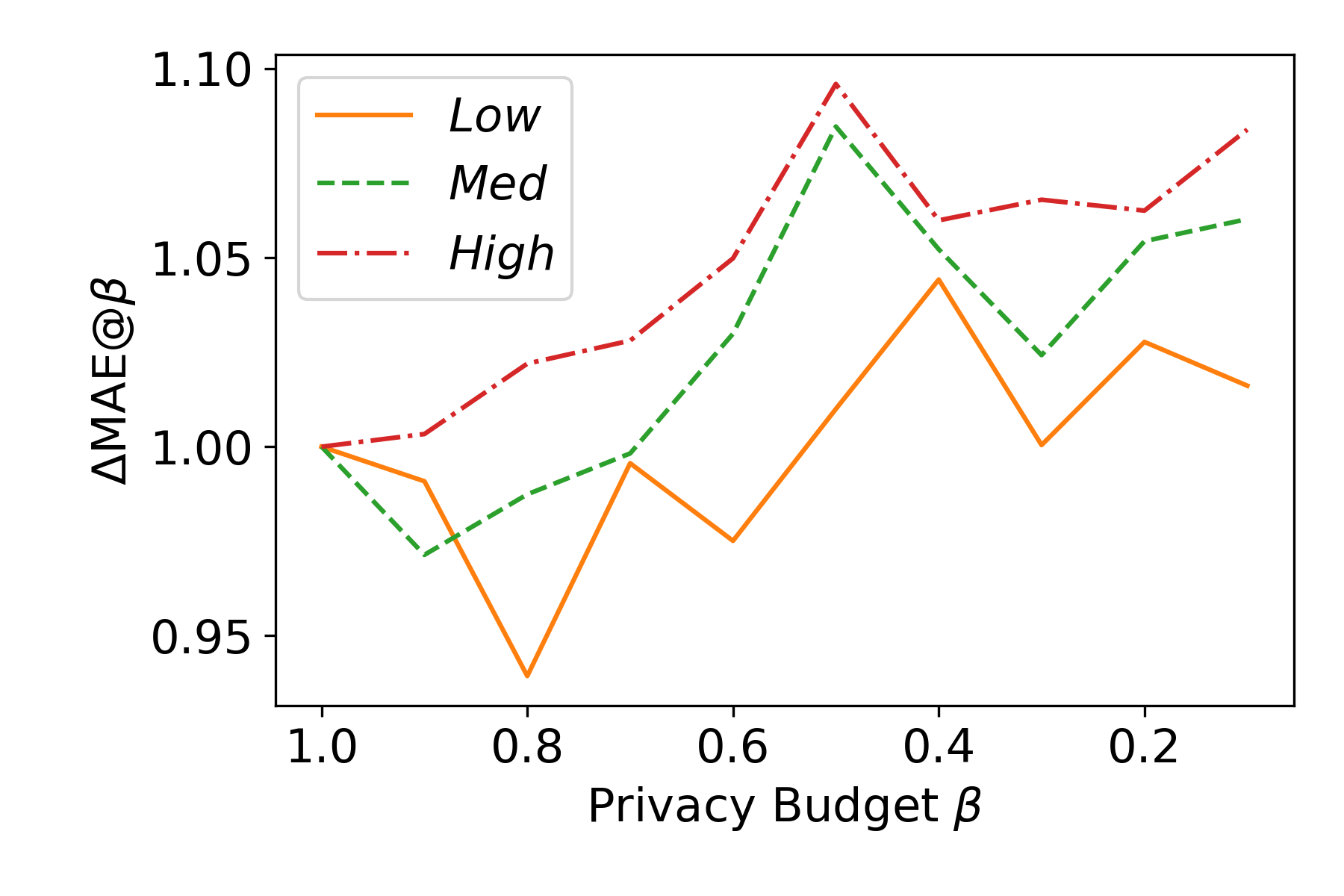}\label{fig:usergroups_d}}
    \subfloat[t][Jester]{\includegraphics[width=0.33\linewidth]{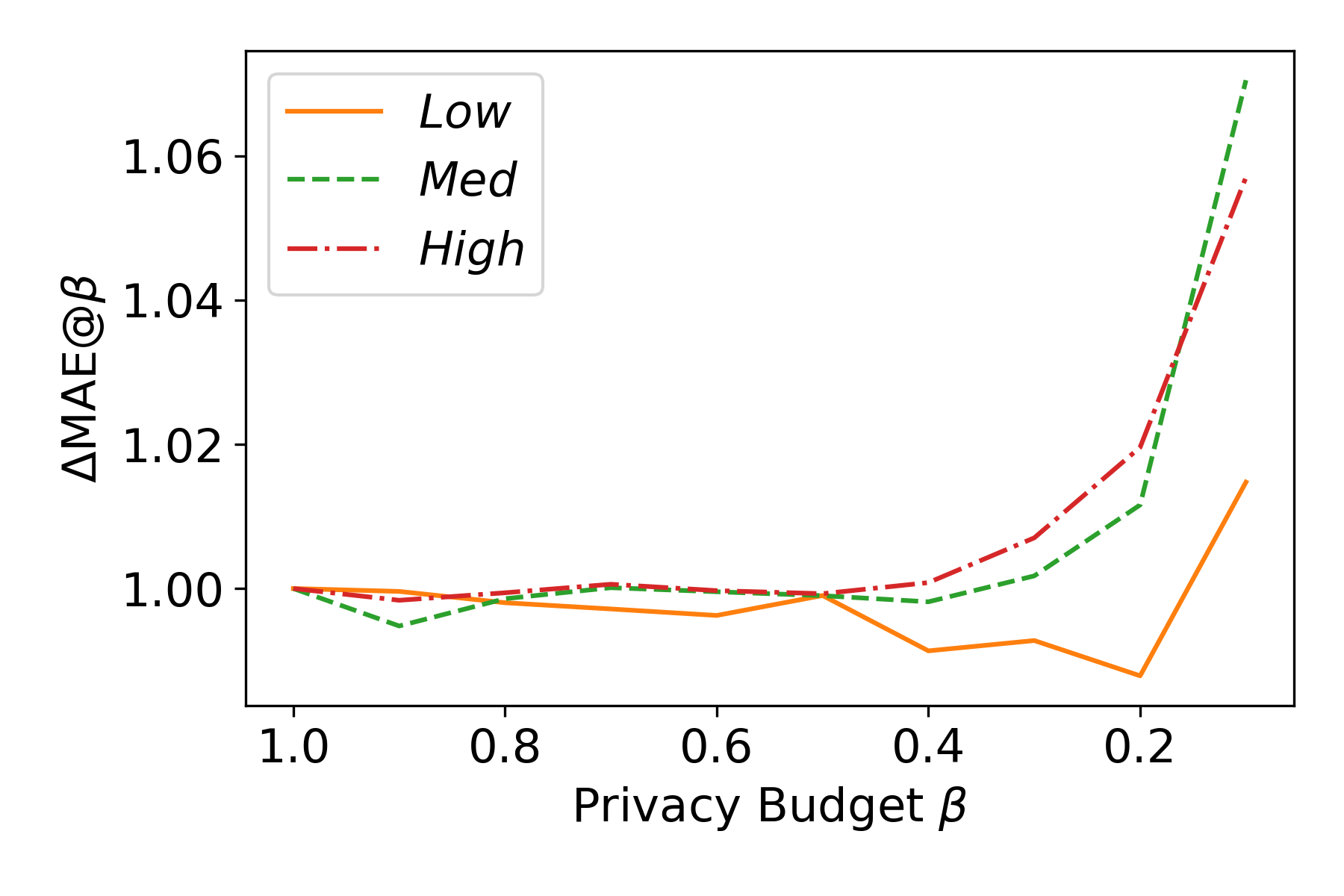}\label{fig:usergroups_e}}
    \end{adjustbox}
    \caption{MetaMF's $\Delta\mathrm{MAE}@\beta$ measurements for the (a) Douban, (b) Hetrec-MovieLens, (c) MovieLens 1M, (d) Ciao, and (e) Jester dataset for all three usergroups. Especially figures (a), (b), and (c) illustrate that \emph{High} is sensitive to small privacy budgets. In contrast, \emph{Low} can afford a high degree of privacy, since the accuracy of its recommendations only marginally decreases (\emph{RQ2b}).}
    \label{fig:usergroups}
\end{figure}

To detail the impact of decreasing privacy budgets on these user groups, we monitor the $\Delta\mathrm{MAE}@\beta$ on \emph{Low}, \emph{Med}, and \emph{High}.
The results for our five datasets are presented in Figure~\ref{fig:usergroups}.
Surprisingly, \emph{Low} seems to be much more robust against small privacy budgets than \emph{High}.
Here, we refer to our observations about MetaMF's performance on the Ciao and Jester dataset in Figure~\ref{fig:meta_nometa_comparison_a}.
In contrast to the other datasets, Ciao and Jester comprise only a small average number of ratings per user, i.e., 38 (Ciao) and 56 (Jester), which means that they share a common property with our \emph{Low} user group.
Thus, we suspect a relationship between the robustness against decreasing privacy budgets and the amount of rating data per user. 
The most prominent examples of \emph{Low} being more robust than \emph{High} can be found in Figures~\ref{fig:usergroups_a}, \ref{fig:usergroups_b} and \ref{fig:usergroups_c}.
Here, the accuracy of MetaMF on \emph{High} substantially decreases for small privacy budgets.
On the one hand, MetaMF provides strongly personalized recommendations for users with lots of ratings, which results in a high accuracy for these users (i.e., \emph{High}).
On the other hand, this personalization leads to a serious reliance on the data, which has a negative impact on the performance in settings with small privacy budgets.
Thus, concerning \emph{RQ2b}, we conclude that users with lots of ratings receive better recommendations than other users if they can take advantage of their abundance of data.
In settings where a high level of privacy is required, i.e., a low privacy budget, and thus, users decide to hold back the majority of their data, users are advantaged who do not require as much personalization from the recommender system.

\section{Conclusions \& Future Work}
In our study at hand, we conducted two lines of research.
First, we reproduced results presented by Lin et al. in~\cite{linmeta}. 
Besides, we introduced a fifth dataset, i.e., Jester, which, in contrast to the originally utilized datasets, has plenty of rating data per item.
We found that all accuracy measurements are indeed reproducible (\emph{RQ1a}).
However, our reproduction of the t-SNE visualizations of the embeddings illustrated potential discrepancies between our and Lin et al.'s work (\emph{RQ1b}).
Second, we conducted privacy-focused studies. 
Here, we thoroughly investigated the meta learning component of MetaMF.
We found that meta learning takes an important role in preserving the accuracy of the recommendations for decreasing privacy budgets (\emph{RQ2a}).
Furthermore, we evaluated MetaMF's performance with respect to decreasing privacy budgets on three user groups that differ in their amounts of rating data. 
Surprisingly, the accuracy of the recommendations for users with lots of ratings seems far more sensitive to small privacy budgets than for users with a limited amount of data (\emph{RQ2b}). 

\subsubsection{Future work.} In our future work, we will research how to cope with incomplete user profiles in our datasets, as users may already have limited the amount of their rating data to satisfy their privacy constraints.
Furthermore, we will develop methods that identify the ratings a user should share based on the characteristics of the data.
\\ \\

\noindent 
\textbf{Acknowledgements.} 
We thank the Social Computing team for their rich feedback on this work. This work is supported by the H2020 project TRUSTS (GA: 871481) and the ``DDAI'' COMET Module within the COMET – Competence Centers for Excellent Technologies Programme, funded by the Austrian Federal Ministry for Transport, Innovation and Technology (bmvit), the Austrian Federal Ministry for Digital and Economic Affairs (bmdw), the Austrian Research Promotion Agency (FFG), the province of Styria (SFG) and partners from industry and academia. The COMET Programme is managed by FFG.

\newpage
\bibliographystyle{splncs04}
\bibliography{bibliography}
\end{document}